\documentclass[sigconf]{acmart}
\usepackage{tabularx}
\usepackage{graphicx}
\usepackage{mathtools}
\usepackage{algorithm2e}
\usepackage{wasysym}
\usepackage{tablefootnote}
\usepackage{multicol}
\usepackage{caption}
\usepackage{subcaption}
\usepackage{textcomp}
\usepackage{import}
\usepackage{xcolor}
\usepackage{etoolbox}
\usepackage{listings}
\usepackage{hyperref}
\usepackage{booktabs}
\usepackage[capitalise,noabbrev]{cleveref}    
\usepackage{pgfplotstable}
\usepgfplotslibrary{groupplots}
\pgfplotsset{compat=1.18}
\usepackage{lipsum}
\usepackage{balance}
\usepackage{import}
\usepackage[inline]{enumitem}
\usepackage{titlesec}
\usepackage{chapterfolder}
\usepackage{comment}
\usepackage{enumitem}
\usepackage{tikz}
\usetikzlibrary{patterns}

\definecolor[named]{ACMBlue}{cmyk}{1,0.1,0,0.1}
\definecolor[named]{ACMYellow}{cmyk}{0,0.16,1,0}
\definecolor[named]{ACMOrange}{cmyk}{0,0.42,1,0.01}
\definecolor[named]{ACMRed}{cmyk}{0,0.90,0.86,0}
\definecolor[named]{ACMLightBlue}{cmyk}{0.49,0.01,0,0}
\definecolor[named]{ACMGreen}{cmyk}{0.20,0,1,0.19}
\definecolor[named]{ACMPurple}{cmyk}{0.55,1,0,0.15}
\definecolor[named]{ACMDarkBlue}{cmyk}{1,0.58,0,0.21}

\definecolor[named]{myGreen}{RGB}{31,182,83}
\definecolor[named]{myPurple}{RGB}{117, 112, 179}
\definecolor[named]{myDarkGreen}{RGB}{27,158,119}
\definecolor[named]{myDarkOrange}{RGB}{217,95,2}

\usepackage[framemethod=tikz]{mdframed}
\mdfdefinestyle{mystyle}{%
  rightline=true,
  innerleftmargin=5,
  innerrightmargin=5,
  outerlinewidth=0.8pt,
  topline=true,
  leftline=true,
  bottomline=true,
  skipabove=\topsep
}

\AtBeginDocument{%
  \providecommand\BibTeX{{%
    \normalfont B\kern-0.5em{\scshape i\kern-0.25em b}\kern-0.8em\TeX}}}

\setcopyright{acmlicensed}
\copyrightyear{2024}
\acmYear{2024}
\acmDOI{XXXXXXX.XXXXXXX}

\acmConference[Conference acronym 'XX]{Make sure to enter the correct
  conference title from your rights confirmation emai}{June 03--05,
  2018}{Woodstock, NY}
\acmISBN{978-1-4503-XXXX-X/18/06}

\begin{document}

\title{Efficient Data Access Paths for Mixed Vector-Relational Search}

\author{Viktor Sanca}
\email{viktor.sanca@epfl.ch}
\orcid{0000-0002-4799-8467}
\affiliation{%
  \institution{EPFL}
  \country{Switzerland}}

% \author{Lubor Budaj}
% %\authornotemark[1]
% \email{lubor.budaj@epfl.ch}
% %\orcid{1234-5678-9012}
% \affiliation{%
%   \institution{EPFL}
%   \country{Switzerland}}

\author{Anastasia Ailamaki}
\email{anastasia.ailamaki@epfl.ch}
\orcid{0000-0002-9949-3639}
\affiliation{%
  \institution{EPFL}
  \country{Switzerland}}

\renewcommand{\shortauthors}{Sanca, et al.}

\begin{abstract}
The rapid growth of machine learning capabilities and the adoption of data processing methods using vector embeddings sparked a great interest in creating systems for vector data management. While the predominant approach of vector data management is to use specialized index structures for fast search over the entirety of the vector embeddings, once combined with other (meta)data, the search queries can also become selective on relational attributes - typical for analytical queries. As using vector indexes differs from traditional relational data access, we revisit and analyze alternative access paths for efficient mixed vector-relational search. 

We first evaluate the accurate but exhaustive scan-based search and propose hardware optimizations and alternative tensor-based formulation and batching to offset the cost. We outline the complex access-path design space, primarily driven by relational selectivity, and the decisions to consider when selecting an exhaustive scan-based search against an approximate index-based approach. Since the vector index primarily avoids expensive computation across the entire dataset, contrary to the common relational knowledge, it is better to scan at lower selectivity and probe at higher, with a cross-point between the two approaches dictated by data dimensionality and the number of concurrent search queries.

%Vector databases typically rely on indexes - overall search where the expression is on vector similarity.
%However, data might also have the factor of relational attributes, and selective on those, but which might not be compatible with the index access cost, where a single index query/probe is in the millisecond range.

%We thus optimize the and improve the scan (in-memory, scale-up), and compare it against concurrent index access, and show that the complex design space and complex access patterns have tradeoffs.
%This is unexpected, as scan is now better for selective queries, as vector indexes partition the space and enable avoiding not only data access, but comparatively expensive computation.

% NEXT: layouts for indexes and vectors (row, pax, column - scan or probe, which index type) 
% NEXT: caching/layouts (avoid computation if it is a join/batch/similar enough vector)
% NEXT: analysis against more indexes and use-cases
% NEXT: scalability analysis + modern hardware (CPU/GPU)
\end{abstract}

\ccsdesc[500]{Information systems~Data access methods}
\ccsdesc[500]{Information systems~Data scans}
\ccsdesc[500]{Information systems~Point lookups}
\ccsdesc[300]{Information systems~Database query processing}
\ccsdesc[100]{Information systems~Parallel and distributed DBMSs}
\ccsdesc[300]{Information systems~Main memory engines}
\ccsdesc[300]{Information systems~Online analytical processing engines}
\ccsdesc[100]{Information systems~Data analytics}
\ccsdesc[300]{Information systems~Retrieval models and ranking}
\ccsdesc[500]{Information systems~Search engine architectures and scalability}

\keywords{Access Methods, Vector Databases, Index Structures, Query Optimization, Vector-Relational Databases}

\received{22 March 2024}

% page numbering
\settopmatter{printfolios=true}

\maketitle

% 6 page submission, 10 page final + references

% 

\section{Introduction}

The widespread surge in the popularity and use of LLMs and embedding-driven ML models resulted in an urgent need for systems that support storing and processing vector embeddings. Use-cases such as Retrieval-Augmented Generation~\cite{DBLP:conf/nips/LewisPPPKGKLYR020}, search~\cite{huang2020embedding}, and caching~\cite{DBLP:conf/cvpr/ZhangHLHDQGL23,DBLP:conf/vldb/SancaA23a} are currently observed and optimized mostly as \textit{vector-only} processing.
Still, the topic of vector index structures representing a fundamental building block of vector management isn't new, being thoroughly studied over the years~\cite{DBLP:journals/pami/MalkovY20, DBLP:conf/osdi/ZhangXCSXCCH00Y23, DBLP:conf/sosp/XuLLXCZLYYYCY23, DBLP:journals/tbd/JohnsonDJ21, rapidsai}.

Efficient data access and processing is fundamental to decades of research and industrial data management systems efforts. This has resulted in studies of data layouts, index structures, and using various storage and processing hardware for optimizing and selecting data access methods~\cite{DBLP:conf/sigmod/KesterAI17}. However, research and industrial efforts typically target operations and the processing of relational data, resulting in benchmarks such as TPC-H, TPC-DS~\cite{tpc2024}, and Star Schema Benchmark~\cite{DBLP:conf/tpctc/ONeilOCR09}.
On the other hand, with the proliferation of use, building efficient vector search systems and components becomes an imperative, resulting in corresponding test suites and evaluations such as ANN-Benchmark~\cite{ANN-Benchmarks, DBLP:journals/is/AumullerBF20}.

As model-driven vector capabilities become commonplace and start interacting with (relational) analytical systems driven by novel use-cases~\cite{DBLP:conf/icde/SancaA23, DBLP:journals/corr/abs-2312-01476}, the mix of both relational and vector data access characteristics will bring a richer design space to consider. Applying both relational filters and vector similarity search is relevant when not all the embeddings should be considered, for example, searching for embeddings in a particular date range or having certain characteristics, such as in document tagging~\cite{DBLP:conf/rep4nlp/ChenSPM17} or misinformation detection~\cite{AIMeta2024CovidMisinformation}. In a simpler case, the user might want to search for images similar to a given one but only taken in a certain location in a given date range, both of which we consider relational data.

We consider the mixed vector-relational data use case and study the data access path characteristics and trade-offs for the in-memory scale-up case and: 
\begin{itemize}
    \item outline the parameters impacting the access path selection for relational, vector, and the mixed selective data access in \cref{sec:background},
    \item describe the landscape of scan-based, exhaustive search strategies and introduce a novel batch-based tensor search and propose hardware optimizations in \cref{sec:scan}, and describe relational filtering approaches for vector indexes in \cref{sec:probe}.
    \item evaluate the different access strategies in \cref{sec:eval}, and demonstrate the parameters affecting the selection of the most efficient strategy, showing that the access path selection needs to be adaptive based on the workload and particular execution parameters such as selectivity. 
\end{itemize}

Using a particular access path can yield orders of magnitude difference, reinforcing the need to adapt the strategy based on the runtime needs rather than having a single static method.

%In contrast to indexes for relational data, vector indexes are more efficient when the selectivity is high, introducing an expected
%We know about the tradeoffs (and give some examples on selectivity)~\cite{DBLP:conf/sigmod/KesterAI17}, and why this is difficult/currently looked into
%Brute force, exhaustive search - if we talk about in-memory or SSDs, there is high bandwidth. Scan vs random access cost.
    % 0.5 p     -- 1
\section{Background}
\label{sec:background}

Selecting a subset of qualifying records to process is a fundamental data processing operation.
Still, there are several differences between the relational and vector search contexts.

\subsection{Traditional relational search}

Relational search most commonly filters out results based on a predicate which is often a numerical expression, and yields an exact result set of data records that satisfy the condition.

\begin{figure}
\centering
    %%%%%%%%%%%%%%%%%%%%%%%%%%%%%%%%%%%%%%%%%%%%%%%%%%%%%%%%%%%%%
        \begin{tikzpicture}

        \begin{axis}[
            legend style={at={(0.5,1)},draw=none,
                anchor=south,
                legend columns=2,
                inner sep=0pt, outer sep=0pt, font=\small%,
            },
            axis lines = left,
            axis x line*=bottom,
            axis y line*=left,
            xlabel={\% selectivity},
            xlabel style={align=center},
            ylabel style={align=center},
            ylabel={Time [ms]},
            % ylabel near ticks,
            % legend style={font=\tiny},
            ylabel absolute, every axis y label/.append style={yshift=-1em, font=\small},
            height=0.6\columnwidth,
            width=\linewidth,
            xmin=-1,
            xmax=100,
            ymode=log,
            %log basis y={10},
            % log ticks with fixed point,
            %line width=1.5pt
            %ymin=0,
            %xmin=1,
            %ymax=18000
        ]
            \addplot[color=ACMDarkBlue, thick]  coordinates {
                (0.01,594.04299075)
                (0.05,581.9721505)
                (0.1,610.65246075)
                (0.5,613.37754825)
                (1,575.75)
                (5,573)
                (10,617.5)
                (25,552.5)
                (50,550.5)
                (75,552)
                (90,555.25)
                (95,553.25)
                (99,597.5)
            };
            \addlegendentry{Scan-1T}
    
        	\addplot[color=ACMDarkBlue, thick, dashed]  coordinates {
        		(0.01,1.076477)
                (0.05,4.76601825)
                (0.1,9.90234675)
                (0.5,58.91007075)
                (1,183.5)
                (5,691)
                (10,1371.75)
                (25,3736.75)
                (50,8025.25)
                (75,12174)
                (90,15371.5)
                (95,16900.5)
                (99,17696.75)
        	};
        	\addlegendentry{Probe-1T}

         \addplot[color=ACMBlue, thick]  coordinates {
                (0.01,61.8179125)
                (0.05,60.359607)
                (0.1,60.497893)
                (0.5,61.18677375)
                (1,63)
                (5,64.25)
                (10,65)
                (25,57.5)
                (50,61.25)
                (75,59.5)
                (90,57)
                (95,52)
                (99,48.5)
            };
            \addlegendentry{Scan-48T}

            \addplot[color=ACMBlue, thick, dashed]  coordinates {
        		(0.01,0.348933)
                (0.05,0.41970925)
                (0.1,0.53526675)
                (0.5,2.22003275)
                (1,10.5)
                (5,56.75)
                (10,110.5)
                (25,298.5)
                (50,621.5)
                (75,947.5)
                (90,1026)
                (95,1030.75)
                (99,1017.5)
        	};
        	\addlegendentry{Probe-48T}
	\end{axis}
    \end{tikzpicture}

%\vspace*{-10pt}
\caption{Scan vs Probe in relational context (SUM query), in-memory, using 1 and 48 threads.}
\label{fig:scan_vs_probe_db}
\vspace*{-10pt}
\end{figure}
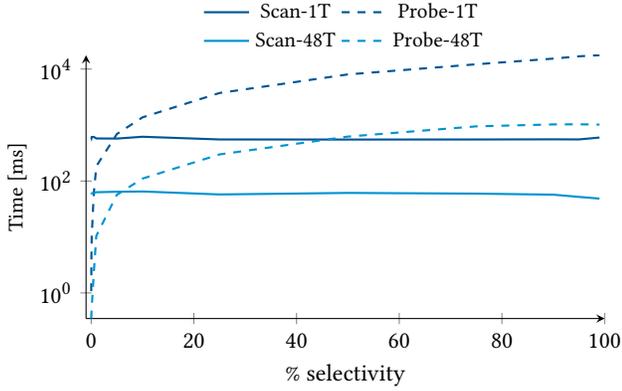

Subject to the nature of the predicate, this typically results in a point search or a range search with varying selectivity. Depending on the sortedness of the data, the presence of auxiliary data structures such as indexes (and their nature, e.g. primary/secondary, hash/tree), and the estimated cost based on selectivity, the query optimizer makes a decision that reduces the access cost.
An index on the search key enables finding the qualifying data but comes at a random access cost and should be used when the selectivity is low. We demonstrate this in a simple in-memory SCAN-FILTER-SUM query (\cref{fig:scan_vs_probe_db}), where a sequential scan with a lightweight filtering predicate processed at memory bandwidth has a better per-tuple throughput. The more cores are allocated, the more per-core memory bandwidth is available (up to the maximum and based on configuration), speeding up the sequential access. Conversely, more cores can issue concurrent requests and probe an index structure. An optimizer should select the access path based primarily on the estimated selectivity of the query.

%primarily random vs sequential data access

\subsection{Vector-based search}
Vector search aims to retrieve other vectors that satisfy the provided distance function or expression in a given range or by retrieving top-k results. This entails comparing high-dimensional vectors, which is computationally intensive and requires accessing and moving high-dimensional tuples across the memory hierarchy. 

To minimize the cost of an exhaustive but accurate search, vector indexes aim to organize and transform the vector data in a way that reduces the search across all the embeddings, based on the construction-time parameters such as the desired distance function~\cite{DBLP:journals/pami/MalkovY20, DBLP:conf/osdi/ZhangXCSXCCH00Y23, DBLP:conf/sosp/XuLLXCZLYYYCY23, DBLP:journals/tbd/JohnsonDJ21, rapidsai}. This way, a more efficient approximate similarity search can take place, especially considering the size of the search space~\cite{DBLP:journals/tbd/JohnsonDJ21}. This search aims to reduce data access and mitigate the significant impact of expensive comparisons between high-dimensional vectors. 

%both computationally heavy and avoiding data access, dimensionality, similarity function

%vector indexes

%relational indexes

%selective queries and access path selection

\paragraph{Takeaway}
While both types of indexes aim to reduce data access and retrieve relevant data records, the per-tuple predicate evaluation is computationally more expensive for vector data.      % 0.5 p     -- 1.5
\section{Scan: Exhaustive Mixed Search}
\label{sec:scan}

We consider a column-store layout that stores both vector and relational data. We start from describing and analyzing scan-based brute-force approach and formalize the selection operator over vector data in the table (relation) $R$ using vector $v$ as the search parameter (query) and $\epsilon$ as distance function:

\begin{equation}
\tag{Selection with query vector $v$ and similarity predicate $\epsilon$}
\sigma_{\epsilon, v}(R) = \{ t \in R,\ \epsilon(t, v) \}
\end{equation}

In a pure vector search, the scan-based approach is equivalent to the cross-product of the search query $v$ and the vector column of table $R$.

\begin{equation}
\tag{Brute force selection.}
\sigma_{\epsilon, v}(R) \Leftrightarrow \sigma_{\epsilon}(v \times R)  
\end{equation}

In the case of a mixed vector-relational search with a relational predicate $\theta_R$, the predicate can be pushed down to the relational columns of table $R$, reducing the number of candidate vectors to compare with.

\begin{equation}
\tag{Brute force similarity search with relational predicate $\theta$.}
\sigma_{\epsilon, v, \theta_{R}}(R) \Leftrightarrow \sigma_{\epsilon, \theta_{R}}(v \times R) \Leftrightarrow \sigma_{\epsilon}(v \times \sigma_{\theta_{R}}(R)) 
\end{equation}

\subsection{Optimization Strategies}

\begin{figure}[h]
  \centering
  \includegraphics[width=0.8\linewidth]{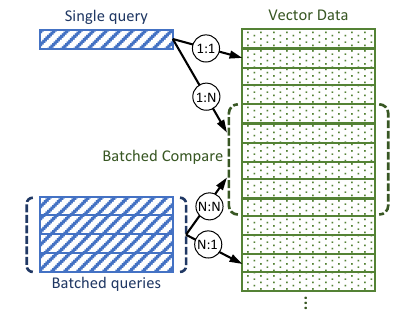}
  \caption{Scan-based vector selection strategies: $1:1$ single query-single data, $1:N$ single query-batched data, $N:N$ batched query-batched data, $N:1$ batched query-single data.}
  \Description{Scan-based vector selection strategies.}
  \label{fig:selection-strategies}
\end{figure}

We observe four possible strategies for scan-based processing (\cref{fig:selection-strategies}). $1:1$ compares a single query vector with a single data vector at a time, where hardware-based optimizations such as using vectorized (SIMD) instructions can speed up the processing of individual vectors. 
An alternative to individual comparisons is batching the vectors to prioritize cache-local computation and avoid the overhead of excessive data movement ($1:N$, $N:1$, $N:N$). 
Finally, applying the general concept of batching several queries where they can benefit from shared data access and computation~\cite{DBLP:journals/pvldb/GiannikisAK12}, we propose batching both the data and query vectors $v$.

This is an especially applicable case if there are many concurrent or temporally-close requests. In this case, there are opportunities to use more efficient algorithms, which we propose as a batched \textbf{Tensor formulation}, based on BLAS linear algebra algorithms~\cite{blackford2002updated}. By blurring the boundary between individual vectors, the tensor-based computation is more cache-efficient, keeping the data in the caches the longest with efficient vector processing procedures.
For example, if the distance function $\epsilon$ is cosine distance, it can be represented as a dot product of two normalized input matrices. 

\subsection{Relational Predicate Pushdown}
We consider two strategies for the relational predicate pushdown. 
One happens on the fly, appropriate for the $1:1$ and $N:1$ approach, where the relational filter is checked, and if the condition is satisfied, the vector comparison takes place. 

\begin{figure}[h]
  \centering
  \includegraphics[width=0.6\linewidth]{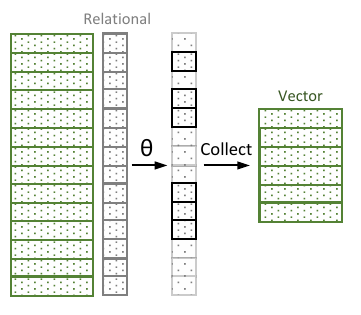}
  \caption{Scan-based vector selection based on a relational predicate-first strategy.}
  \Description{Scan-based vector selection based on a relational predicate.}
  \label{fig:scan-predicate}
\end{figure}

For the batched approach, we propose a late materialization-based~\cite{abadi2006materialization} strategy by first staging the fast relational predicate filtering, then collecting the vectors that satisfy the condition, which are then processed in a batch using the \textbf{Tensor approach} (\cref{fig:scan-predicate}).

\paragraph{Takeaway} Scan-based exhaustive approaches offer sequential data access and optimization opportunities that can benefit further from optimizing the data access to the memory hierarchy, such as batching. This approach excels in relational filtering, typically filtering values at memory scan speed, and is expected to bring benefits with highly selective predicates and when the search accuracy (recall) is prioritized.

%\newpage            % 1         -- 2.5
\section{Index: Probe-Based (Approximate) Search}
\label{sec:probe}

Index-based search relies on constructing data structures and organizing data such that similar data are clustered together. One such approach is Hierarchical Navigable Small World (HNSW)~\cite{DBLP:journals/pami/MalkovY20}, which is an algorithm based on the idea of \textit{small-world networks} where most nodes can be reached from any other by a small number of steps, despite the large initial network. This results in a multi-layered graph-based structure, where each layer is a (probabilistic) subset of the layer below, where the bottom layer contains all the data points. The probabilistic search procedure provides an \textbf{approximate} nearest neighbor search capability, tunable to the recall and speed properties.

\subsection{Relational Filtering}

Current vector indexes typically reorganize the underlying data and don't natively support relational filtering~\cite{DBLP:conf/osdi/ZhangXCSXCCH00Y23}. Still, there are existing approaches to achieve this. 
One approach is to perform the vector search and then apply the relational filter in a \textbf{post-filtering} step. This might invalidate all the results where qualifying tuples aren't immediately the most similar to the query vector.
Another approach is \textbf{pre-filtering}, where the tuples are first checked for the relational condition. To preserve the properties of the data structure and the connectivity of HNSW, the tuples are not discarded for search but simply not included in the result set, influencing the probing of the index.

\paragraph{Takeaway}
Indexes represent a probe-based approach, facing the penalty of random accesses and less-suitable relational filtering. They are more expensive than relational indexes, currently taking milliseconds per vector query for in-memory execution~\cite{ANN-Benchmarks}. They are expected to excel when most of the data is considered for the search, with a crossing point where the scan-based approach computation matches the index-based approach's random access (and the computation). 

%\clearpage          % 0.5 p      -- 3
\section{Evaluation}
\label{sec:eval}

We start by evaluating the optimization of the exhaustive scan-based approach and comparing them against different batch sizes and data dimensionalities, concluding on the best strategy given the parameters.
Then, we focus on selective data access with a scan-based approach.
Finally, we compare the index to the scan-based approach, providing an answer to the question \textit{"Should I Scan or Should I Probe?"\cite{DBLP:conf/sigmod/KesterAI17}} in mixed vector-relational setting with a selectivity-driven analysis.

%We start by demonstrating the functionality of using models as a driver of context-enhanced relational operations through the example of word embeddings. 
%We then focus on the main performance evaluation of the proposed logical and physical optimizations, showing that a holistic approach is necessary to obtain a performant join algorithm.

\paragraph{System} To conduct the in-depth study, we compare all the approaches in a standalone system in C++ and use Intel AVX instructions for SIMD execution. Tensor formulation benchmarks use Intel oneAPI Math Kernel Library for CPU-aware and efficient BLAS-based linear algebra operations. If otherwise not noted, the experiments run on 48 all threads (physical+hyperthreads).

\paragraph{Vector Index} We use \texttt{hnswlib}\cite{DBLP:journals/pami/MalkovY20, hnswlib} as a vector index, which comes with pre-filtering support, and do not account for index construction cost in our experiments, assuming a correct index already exists. We build an index with parameters $M = 64$, $ef\_construction = 512$, and the inner product as the distance function. We use $k = 64$ for top-k retrieval. 

\paragraph{Hardware Setup} We run the end-to-end and scalability experiments on a two-socket Intel Xeon Gold 5118 CPU (2 x 12-core, 48 threads) and 384GB of RAM.

\paragraph{Dataset} All experiments are with in-memory data; experiments with synthetic data use the same random number generator seed for reproducibility. The vector data of various dimensionalities and relational data are generated uniformly at random. If otherwise not noted, we use 1M data points as the dataset and repeat the experiments 5 times with warm caches. We use cosine similarity $>0.9$ as a vector distance metric.

%%%%%%%%%%%%%%%%%%%%%%%%%%%%%%%%%%%%%%%%
\subsection{Which scan-based approach to select?}

%\input{05-evaluation/scan_optimization_micro}

% We first compare two strategies, running the fully optimized NLJ with hardware vectorization against the Tensor formulation. 
% For this, we vary two factors: the total number of floating point numbers processed (\#FP32 Ops) and how many floating point numbers represent an individual vector (vector dimensionality, Vector \#FP32).
% %\cref{fig:micro_approaches} summarizes the findings, where three data clusters are based on the number of operations, refined by individual vector size. 
% We use the per-FP32 breakdown as a unifying metric across the input size and dimensionality.
% First, we notice the benefit of vectorization with increased vector size, where specialized hardware operations improve the per-tuple performance.
% Second, pushing this boundary beyond per-tuple-vector but to a whole tuple-vector-batch (Tensor), when sufficient computation can benefit from the cache locality, significantly improves the execution time in the batched case

\begin{figure}
    %\color{ACMBlue} % for revision
    \centering
    
    \begin{subfigure}{\linewidth}
    \centering
    %%%%%%%%%%%%%%%%%%%%%%%%%%%%%%%%%%%%%%%%%%%%%%%%%%%%%%%%%%%%%
        \begin{tikzpicture}

        \begin{axis}[
            legend style={at={(0.5,1.1)},draw=none,
                anchor=south,
                legend columns=3,%-1,
                %\addlegendentry{\hspace{-.6cm}\textbf{A title}},
                % /tikz/every even column/.append style={column sep=0.075\textwidth},
                inner sep=0pt, outer sep=0pt%,
                % legend columns=3
            },
            %y filter/.code=\pgfmathparse{#1/1000}, % file contains ms
            % xtick=data,
            axis lines = left,
            axis x line*=bottom,
            axis y line*=left,
            xlabel={Tuples in a query batch},
            xlabel style={align=center},
            ylabel style={align=center},
            ylabel={Time [ms]},
            % ylabel near ticks,
            % legend style={font=\tiny},
            ylabel absolute, every axis y label/.append style={yshift=-1em},
            enlarge x limits=0.25,
            height=0.4\columnwidth,
            width=\linewidth,
            % ymin=0.0005,
            % xmode=log,
            % ymax=30,
            % xmin=0.002,
            ymode=log,
            log basis y={10},
            % log ticks with fixed point,
            %line width=1.5pt
            ymin=1,
            xmin=1,
            xtick  = {1,2,3},
            xticklabels={
                1,
                100,
                10k,
            },
            bar width=10pt,
            %log x ticks with fixed point/.style={
            %xticklabel={%
            %     \pgfkeys{/pgf/fpu=true}%
            %     \pgfmathparse{(10^\tick)*100}%
            %    \pgfmathprintnumber[fixed relative, precision=3]{\pgfmathresult}\%%
            %}
            %}
            x tick label style={align=center},
            point meta=rawy,
            ymax=400000,
            ybar,
            %nodes near coords,
            %every node near coord/.append style={font=\tiny, rotate=90, anchor=west} % has to be after ybar
        ]
        \addplot[fill=ACMDarkBlue, text=ACMDarkBlue, fill opacity=1, pattern=crosshatch dots, pattern color=ACMDarkBlue]
        	coordinates {(1,5.25) (2,199) (3,19874.5)};
        
        \addplot[fill=ACMDarkBlue, ACMDarkBlue, fill opacity=1, pattern=north east lines, pattern color=ACMDarkBlue]
        	coordinates {(1,6) (2,216.75) (3,15957.75)};
        
        \addplot[fill=myGreen, text=ACMDarkBlue, pattern=horizontal lines, pattern color=myGreen]
        	coordinates {(1,51) (2,93) (3,1354)};

        % \addplot[fill=ACMDarkBlue, text=ACMDarkBlue, fill opacity=1]
        % 	coordinates {(1,4) (2,35) (3,269.6)};
        
        \legend{Function, Explicit SIMD, Tensor}
        \end{axis}
        \end{tikzpicture}

    %\vspace*{-10pt}
    \caption{Vector dimensionality D=64.}
    \label{fig:scan_comparison_dims_64}
    %\vspace{-10pt}
    \end{subfigure}
    
    \begin{subfigure}{\linewidth}
    \centering
    %%%%%%%%%%%%%%%%%%%%%%%%%%%%%%%%%%%%%%%%%%%%%%%%%%%%%%%%%%%%%
        \begin{tikzpicture}

        \begin{axis}[
            legend style={at={(0.5,1.1)},draw=none,
                anchor=south,
                legend columns=3,%-1,
                %\addlegendentry{\hspace{-.6cm}\textbf{A title}},
                % /tikz/every even column/.append style={column sep=0.075\textwidth},
                inner sep=0pt, outer sep=0pt%,
                % legend columns=3
            },
            %y filter/.code=\pgfmathparse{#1/1000}, % file contains ms
            % xtick=data,
            axis lines = left,
            axis x line*=bottom,
            axis y line*=left,
            xlabel={Tuples in a query batch},
            xlabel style={align=center},
            ylabel style={align=center},
            ylabel={Time [ms]},
            % ylabel near ticks,
            % legend style={font=\tiny},
            ylabel absolute, every axis y label/.append style={yshift=-1em},
            enlarge x limits=0.25,
            height=0.4\columnwidth,
            width=\linewidth,
            % ymin=0.0005,
            % xmode=log,
            % ymax=30,
            % xmin=0.002,
            ymode=log,
            log basis y={10},
            % log ticks with fixed point,
            %line width=1.5pt
            ymin=1,
            xmin=1,
            xtick  = {1,2,3},
            xticklabels={
                1,
                100,
                10k,
            },
            bar width=10pt,
            %log x ticks with fixed point/.style={
            %xticklabel={%
            %     \pgfkeys{/pgf/fpu=true}%
            %     \pgfmathparse{(10^\tick)*100}%
            %    \pgfmathprintnumber[fixed relative, precision=3]{\pgfmathresult}\%%
            %}
            %}
            x tick label style={align=center},
            point meta=rawy,
            ymax=400000,
            ybar,
            %nodes near coords,
            %every node near coord/.append style={font=\tiny, rotate=90, anchor=west} % has to be after ybar
        ]
        \addplot[fill=ACMDarkBlue, text=ACMDarkBlue, fill opacity=1, pattern=crosshatch dots, pattern color=ACMDarkBlue]
        	coordinates {(1,68.75) (2,2463) (3,263414.5)};
        
        \addplot[fill=ACMDarkBlue, ACMDarkBlue, fill opacity=1, pattern=north east lines, pattern color=ACMDarkBlue]
        	coordinates {(1,68) (2,603.25) (3,128401.75)};
        
        \addplot[fill=myGreen, text=ACMDarkBlue, pattern=horizontal lines, pattern color=myGreen]
        	coordinates {(1,550) (2,870) (3,17867.75)};

        % \addplot[fill=ACMDarkBlue, text=ACMDarkBlue, fill opacity=1]
        % 	coordinates {(1,4) (2,35) (3,269.6)};
        \end{axis}
        \end{tikzpicture}

    %\vspace*{-10pt}
    \caption{Vector dimensionality D=1024.}
    \label{fig:scan_comparison_dims_1024}
    \vspace{-10pt}
    \end{subfigure}

    \caption{Different scan strategies for 1M tuples and dimensionalities.}
    \label{fig:scan_comparison_dims_vec}
    %\vspace*{-10pt}
\end{figure}
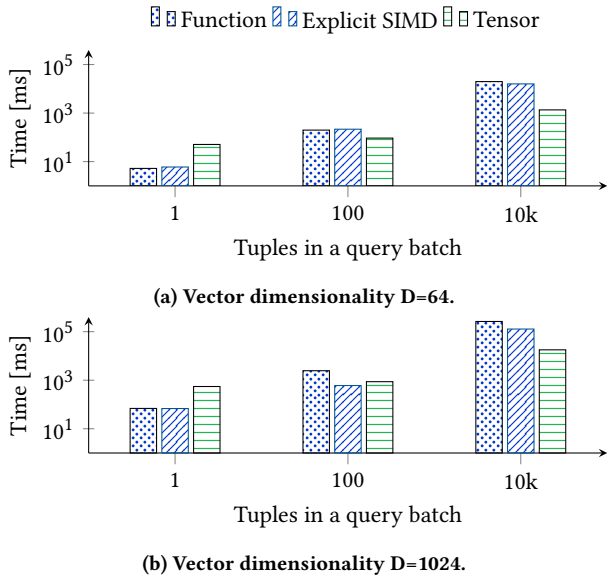

We first compare three strategies, running the per-tuple search with explicit hardware vectorization (\texttt{Explicit SIMD}), inlined function call (\texttt{Function}), and the Tensor formulation (\texttt{Tensor}).
For this, we vary the vector dimensionality and the batch size (\cref{fig:scan_comparison_dims_vec}).

First, we notice the benefit of explicit vectorization with increased vector dimensionality (\cref{fig:scan_comparison_dims_1024}) and in bigger batches where computation dominates. Not using the hardware capabilities is detrimental when computation dominates but less apparent otherwise.
With fewer computations, particularly in small batches and low dimensionality, the tensor-based approach overhead is significant compared to the per-tuple approaches (\cref{fig:scan_comparison_dims_64}). 
However, pushing this boundary beyond per-tuple-vector but to a whole tuple-vector-batch (Tensor), when sufficient computation can benefit from the cache locality, significantly improves the execution time in the high-dimension, large batch case.

\begin{figure}
    %\color{ACMBlue} % for revision
    \centering
    
    \begin{subfigure}{\linewidth}
    \centering
    %%%%%%%%%%%%%%%%%%%%%%%%%%%%%%%%%%%%%%%%%%%%%%%%%%%%%%%%%%%%%
        \begin{tikzpicture}

        \begin{axis}[
            legend style={at={(0.5,1.1)},draw=none,
                anchor=south,
                legend columns=3,%-1,
                %\addlegendentry{\hspace{-.6cm}\textbf{A title}},
                % /tikz/every even column/.append style={column sep=0.075\textwidth},
                inner sep=0pt, outer sep=0pt%,
                % legend columns=3
            },
            %y filter/.code=\pgfmathparse{#1/1000}, % file contains ms
            % xtick=data,
            axis lines = left,
            axis x line*=bottom,
            axis y line*=left,
            xlabel={Vector dimensionality},
            xlabel style={align=center},
            ylabel style={align=center},
            ylabel={Time [ms]},
            % ylabel near ticks,
            % legend style={font=\tiny},
            ylabel absolute, every axis y label/.append style={yshift=-1em},
            enlarge x limits=0.25,
            height=0.4\columnwidth,
            width=\linewidth,
            % ymin=0.0005,
            % xmode=log,
            % ymax=30,
            % xmin=0.002,
            ymode=log,
            log basis y={10},
            % log ticks with fixed point,
            %line width=1.5pt
            ymin=1,
            xmin=1,
            xtick  = {1,2,3},
            xticklabels={
                64,
                256,
                1024,
            },
            bar width=10pt,
            %log x ticks with fixed point/.style={
            %xticklabel={%
            %     \pgfkeys{/pgf/fpu=true}%
            %     \pgfmathparse{(10^\tick)*100}%
            %    \pgfmathprintnumber[fixed relative, precision=3]{\pgfmathresult}\%%
            %}
            %}
            x tick label style={align=center},
            point meta=rawy,
            ybar,
            %nodes near coords,
            %every node near coord/.append style={font=\tiny, rotate=90, anchor=west} % has to be after ybar
        ]
        \addplot[fill=ACMDarkBlue, text=ACMDarkBlue, fill opacity=1, pattern=crosshatch dots, pattern color=ACMDarkBlue]
        	coordinates {(1,5.25) (2,18) (3,68.75)};
        
        \addplot[fill=ACMDarkBlue, ACMDarkBlue, fill opacity=1, pattern=north east lines, pattern color=ACMDarkBlue]
        	coordinates {(1,6) (2,17.75) (3,68)};
        
        \addplot[fill=myGreen, text=ACMDarkBlue, pattern=horizontal lines, pattern color=myGreen]
        	coordinates {(1,51) (2,132) (3,550)};

        % \addplot[fill=ACMDarkBlue, text=ACMDarkBlue, fill opacity=1]
        % 	coordinates {(1,4) (2,35) (3,269.6)};
        
        \legend{Function, Explicit SIMD, Tensor}
        \end{axis}
        \end{tikzpicture}

    \vspace*{-10pt}
    \caption{Query batch size 1 (single query).}
    \label{fig:scan_comparison_dims_1}
    %\vspace{-10pt}
    \end{subfigure}

    \begin{subfigure}{\linewidth}
    \centering
    %%%%%%%%%%%%%%%%%%%%%%%%%%%%%%%%%%%%%%%%%%%%%%%%%%%%%%%%%%%%%
        \begin{tikzpicture}

        \begin{axis}[
            %y filter/.code=\pgfmathparse{#1/1000}, % file contains ms
            % xtick=data,
            axis lines = left,
            axis x line*=bottom,
            axis y line*=left,
            xlabel={Vector dimensionality},
            xlabel style={align=center},
            ylabel style={align=center},
            ylabel={Time [ms]},
            % ylabel near ticks,
            % legend style={font=\tiny},
            ylabel absolute, every axis y label/.append style={yshift=-1em},
            enlarge x limits=0.25,
            height=0.4\columnwidth,
            width=\linewidth,
            % ymin=0.0005,
            % xmode=log,
            % ymax=30,
            % xmin=0.002,
            ymode=log,
            log basis y={10},
            % log ticks with fixed point,
            %line width=1.5pt
            ymin=1,
            xmin=1,
            xtick  = {1,2,3},
            xticklabels={
                64,
                256,
                1024,
            },
            bar width=10pt,
            %log x ticks with fixed point/.style={
            %xticklabel={%
            %     \pgfkeys{/pgf/fpu=true}%
            %     \pgfmathparse{(10^\tick)*100}%
            %    \pgfmathprintnumber[fixed relative, precision=3]{\pgfmathresult}\%%
            %}
            %}
            x tick label style={align=center},
            point meta=rawy,
            ybar,
            %nodes near coords,
            %every node near coord/.append style={font=\tiny, rotate=90, anchor=west} % has to be after ybar
        ]
        \addplot[fill=ACMDarkBlue, text=ACMDarkBlue, fill opacity=1, pattern=crosshatch dots, pattern color=ACMDarkBlue]
        	coordinates {(1,199) (2,659.75) (3,2463)};
        
        \addplot[fill=ACMDarkBlue, ACMDarkBlue, fill opacity=1, pattern=north east lines, pattern color=ACMDarkBlue]
        	coordinates {(1,216.75) (2,213.75) (3,603.25)};
        
        \addplot[fill=myGreen, text=ACMDarkBlue, pattern=horizontal lines, pattern color=myGreen]
        	coordinates {(1,93) (2,204.25) (3,870.75)};

        % \addplot[fill=ACMDarkBlue, text=ACMDarkBlue, fill opacity=1]
        % 	coordinates {(1,4) (2,35) (3,269.6)};
        \end{axis}
        \end{tikzpicture}

    \vspace*{-10pt}
    \caption{Query batch size 100.}
    \label{fig:scan_comparison_dims_100}
    %\vspace{-10pt}
    \end{subfigure}
    
    \begin{subfigure}{\linewidth}
    \centering
    %%%%%%%%%%%%%%%%%%%%%%%%%%%%%%%%%%%%%%%%%%%%%%%%%%%%%%%%%%%%%
        \begin{tikzpicture}

        \begin{axis}[
            %y filter/.code=\pgfmathparse{#1/1000}, % file contains ms
            % xtick=data,
            axis lines = left,
            axis x line*=bottom,
            axis y line*=left,
            xlabel={Vector dimensionality},
            xlabel style={align=center},
            ylabel style={align=center},
            ylabel={Time [ms]},
            % ylabel near ticks,
            % legend style={font=\tiny},
            ylabel absolute, every axis y label/.append style={yshift=-1em},
            enlarge x limits=0.25,
            height=0.4\columnwidth,
            width=\linewidth,
            % ymin=0.0005,
            % xmode=log,
            % ymax=30,
            % xmin=0.002,
            ymode=log,
            log basis y={10},
            % log ticks with fixed point,
            %line width=1.5pt
            ymin=1,
            xmin=1,
            xtick  = {1,2,3},
            xticklabels={
                64,
                256,
                1024,
            },
            bar width=10pt,
            %log x ticks with fixed point/.style={
            %xticklabel={%
            %     \pgfkeys{/pgf/fpu=true}%
            %     \pgfmathparse{(10^\tick)*100}%
            %    \pgfmathprintnumber[fixed relative, precision=3]{\pgfmathresult}\%%
            %}
            %}
            x tick label style={align=center},
            point meta=rawy,
            ybar,
            %nodes near coords,
            %every node near coord/.append style={font=\tiny, rotate=90, anchor=west} % has to be after ybar
        ]
        \addplot[fill=ACMDarkBlue, text=ACMDarkBlue, fill opacity=1, pattern=crosshatch dots, pattern color=ACMDarkBlue]
        	coordinates {(1,19874.5) (2,65109.75) (3,263414.5)};
        
        \addplot[fill=ACMDarkBlue, ACMDarkBlue, fill opacity=1, pattern=north east lines, pattern color=ACMDarkBlue]
        	coordinates {(1,15957.75) (2,24318) (3,128401.75)};
        
        \addplot[fill=myGreen, text=ACMDarkBlue, pattern=horizontal lines, pattern color=myGreen]
        	coordinates {(1,1354) (2,4366) (3,17867.75)};

        % \addplot[fill=ACMDarkBlue, text=ACMDarkBlue, fill opacity=1]
        % 	coordinates {(1,4) (2,35) (3,269.6)};
        \end{axis}
        \end{tikzpicture}

    \vspace*{-10pt}
    \caption{Query batch size 10k.}
    \label{fig:scan_comparison_dims_10k}
    \vspace{-10pt}
    \end{subfigure}

    \caption{Scan strategies, 1M tuples, varying query batch size.}
    \label{fig:scan_comparison_dims}
    %\vspace*{-10pt}
\end{figure}
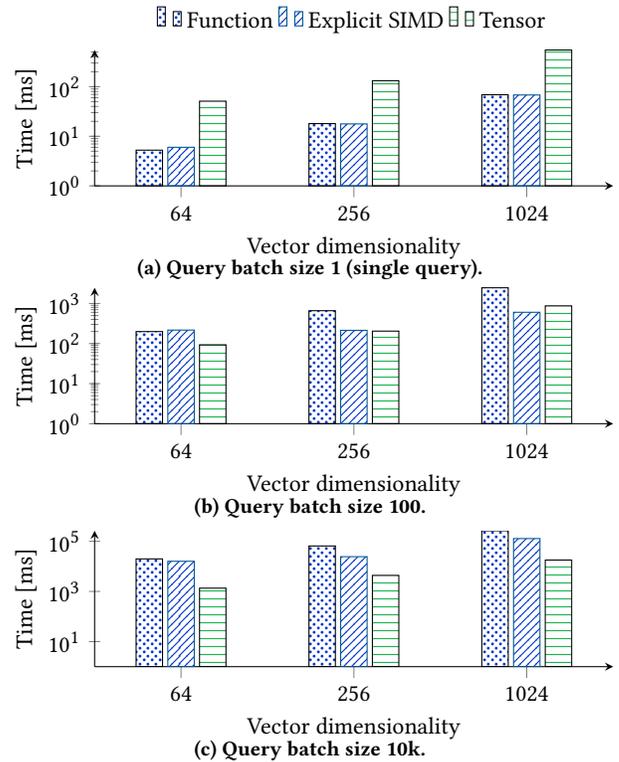

This is more apparent in the per-dimension breakdown (\cref{fig:scan_comparison_dims})
For per-vector search queries, it is best to optimize fully the individual probe, as there is only a performance penalty for setting up the BLAS procedure call (\cref{fig:scan_comparison_dims_1}). 
The more sharing opportunities via batching exist, the more the initial Tensor overhead pays off in larger batches (\cref{fig:scan_comparison_dims_100}, \cref{fig:scan_comparison_dims_10k}), resulting in almost an order of magnitude difference in execution time between the Tensor and Explicit SIMD approach in the case of batch size 10k.

\paragraph{Takeaway} A hardware-optimized, non-batched search is more efficient for individual vector queries. For larger query batches, the tensor-based approach enables faster computation over many queries due to a more efficient, cache-conscious BLAS algorithm.

%%%%%%%%%%%%%%%%%%%%%%%%%%%%%%%%%%%%%%%%
\subsection{Selective exhaustive search}
We now address whether relational selectivity impacts the previous findings on batching and selecting the scan-based approach. To answer this, we show the time breakdown of various steps of \texttt{Tensor} approach, as also outlined in \cref{fig:scan-predicate}. First, the relational filtering happens, then the vector data is (re-)materialized into dense vectors, then normalized (for cosine similarity computation), and then the dot product is computed.

% \begin{equation}
% \tag{Cosine Similarity}
% \label{eq:cosine-sim}
% cos(\theta) = \frac{A \cdot B}{\lVert A \rVert\lVert B \rVert} = \frac{\sum\limits_{i=1}^{n}A_i B_i}{\sqrt{\sum\limits_{i=1}^{n}A_i^{2}}\sqrt{\sum\limits_{i=1}^{n}B_i^{2}}}
% \end{equation}
% \vspace{-15pt}

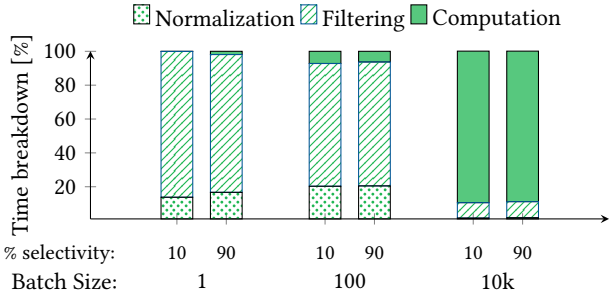
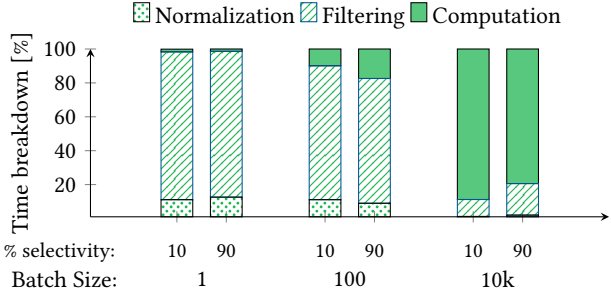
\begin{figure}
    %\color{ACMBlue} % for revision
    \centering
    
    \begin{subfigure}{\linewidth}
    \centering
    %%%%%%%%%%%%%%%%%%%%%%%%%%%%%%%%%%%%%%%%%%%%%%%%%%%%%%%%%%%%%
        \begin{tikzpicture}

        \begin{axis}[
            legend style={at={(0.5,1.1)},draw=none,
                anchor=south,
                legend columns=3,%-1,
                %\addlegendentry{\hspace{-.6cm}\textbf{A title}},
                % /tikz/every even column/.append style={column sep=0.075\textwidth},
                inner sep=0pt, outer sep=0pt%,
                % legend columns=3
            },
            %y filter/.code=\pgfmathparse{#1/1000}, % file contains ms
            % xtick=data,
            axis lines = left,
            axis x line*=bottom,
            axis y line*=left,
            xlabel style={align=center},
            ylabel style={align=center},
            ylabel={Time breakdown [\%]},
            % ylabel near ticks,
            % legend style={font=\tiny},
            ylabel absolute, every axis y label/.append style={yshift=-1em},
            enlarge x limits=0.25,
            height=0.45\columnwidth,
            width=\linewidth,
            % ymin=0.0005,
            % xmode=log,
            % ymax=30,
            % xmin=0.002,
            %ymode=log,
            %log basis y={10},
            % log ticks with fixed point,
            %line width=1.5pt
            ymin=1,
            xmin=1,
            xtick  = {1,2,4,5,7,8},
            xticklabels={
                10,
                90,
                10,
                90,
                10,
                90
            },
            x tick label style={font=\small,align=center,yshift=-5pt},
            point meta=rawy,
            %ymax=15,
            %nodes near coords,
            %every node near coord/.append style={font=\tiny},
            clip=false,
            bar width=12pt,
            %log x ticks with fixed point/.style={
            %xticklabel={%
            %     \pgfkeys{/pgf/fpu=true}%
            %     \pgfmathparse{(10^\tick)*100}%
            %    \pgfmathprintnumber[fixed relative, precision=3]{\pgfmathresult}\%%
            %}
            %}
            x tick label style={align=center},
            point meta=rawy,
            ybar stacked,
            %nodes near coords,
            %every node near coord/.append style={rotate=15} % has to be after ybar
        ]
        \addplot[fill=ACMDarkBlue, text=ACMDarkBlue, fill opacity=1, pattern=crosshatch dots, pattern color=myGreen]
        	%coordinates {(1,48) (2,560) (4,53) (5,462) (7,46) (8,415)}; % derive percentages
            coordinates {(1,13.8) (2,16.7) (4,20.3) (5,20.5) (7,1.5) (8,1.7)};
        
        \addplot[fill=ACMDarkBlue, ACMDarkBlue, fill opacity=1, pattern=north east lines, pattern color=myGreen]
        	%coordinates {(1,376.34) (2,3765.34) (4,376.34) (5,3765.34) (7,376.34) (8,3765.34)};
            coordinates {(1,86.1) (2,81.4) (4,72.5) (5,73.2) (7,9.1) (8,9.5)};
        
        \addplot[fill=myGreen, fill opacity=0.75, text=ACMDarkBlue, pattern color=myGreen]
        	%coordinates {(1,7) (2,62.5) (4,47) (5,890) (7,3327) (8,16091.75)};
            coordinates {(1,0.1) (2,1.9) (4,7.2) (5,6.3) (7,89.5) (8,88.9)};

        % \addplot[fill=ACMDarkBlue, text=ACMDarkBlue, fill opacity=1]
        % 	coordinates {(1,4) (2,35) (3,269.6)};

            \node[] at (xticklabel cs:.2150,5pt) {1};
            \node[] at (xticklabel cs:.5000,5pt) {100};
            \node[] at (xticklabel cs:.7850,5pt) {10k};
    
            \node[] at (xticklabel cs:-0.05,5pt) {Batch Size:};
            \node[font=\small] at (xticklabel cs:-0.06,-5pt) {\% selectivity:};

        \legend{Normalization, Filtering, Computation}
        \end{axis}
        \end{tikzpicture}

    %\vspace*{-10pt}
    \caption{Vector dimensionality D=64.}
    \label{fig:gemm_breakdown_64}
    %\vspace{-10pt}
    \end{subfigure}
    
    \begin{subfigure}{\linewidth}
    \centering
    %%%%%%%%%%%%%%%%%%%%%%%%%%%%%%%%%%%%%%%%%%%%%%%%%%%%%%%%%%%%%
        \begin{tikzpicture}

        \begin{axis}[
            legend style={at={(0.5,1.1)},draw=none,
                anchor=south,
                legend columns=3,%-1,
                %\addlegendentry{\hspace{-.6cm}\textbf{A title}},
                % /tikz/every even column/.append style={column sep=0.075\textwidth},
                inner sep=0pt, outer sep=0pt%,
                % legend columns=3
            },
            %y filter/.code=\pgfmathparse{#1/1000}, % file contains ms
            % xtick=data,
            axis lines = left,
            axis x line*=bottom,
            axis y line*=left,
            xlabel style={align=center},
            ylabel style={align=center},
            ylabel={Time breakdown [\%]},
            % ylabel near ticks,
            % legend style={font=\tiny},
            ylabel absolute, every axis y label/.append style={yshift=-1em},
            enlarge x limits=0.25,
            height=0.45\columnwidth,
            width=\linewidth,
            % ymin=0.0005,
            % xmode=log,
            % ymax=30,
            % xmin=0.002,
            %ymode=log,
            %log basis y={10},
            % log ticks with fixed point,
            %line width=1.5pt
            ymin=1,
            xmin=1,
            xtick  = {1,2,4,5,7,8},
            xticklabels={
                10,
                90,
                10,
                90,
                10,
                90
            },
            x tick label style={font=\small,align=center,yshift=-5pt},
            point meta=rawy,
            %ymax=15,
            %nodes near coords,
            %every node near coord/.append style={font=\tiny},
            clip=false,
            bar width=12pt,
            %log x ticks with fixed point/.style={
            %xticklabel={%
            %     \pgfkeys{/pgf/fpu=true}%
            %     \pgfmathparse{(10^\tick)*100}%
            %    \pgfmathprintnumber[fixed relative, precision=3]{\pgfmathresult}\%%
            %}
            %}
            x tick label style={align=center},
            point meta=rawy,
            ybar stacked,
            %nodes near coords,
            %every node near coord/.append style={rotate=15} % has to be after ybar
        ]
        \addplot[fill=ACMDarkBlue, text=ACMDarkBlue, fill opacity=1, pattern=crosshatch dots, pattern color=myGreen]
        	%coordinates {(1,48) (2,560) (4,53) (5,462) (7,46) (8,415)}; % derive percentages
            coordinates {(1,11.1) (2,12.6) (4,11.1) (5,9.0) (7,1.2) (8,2.0)};
        
        \addplot[fill=ACMDarkBlue, ACMDarkBlue, fill opacity=1, pattern=north east lines, pattern color=myGreen]
        	%coordinates {(1,376.34) (2,3765.34) (4,376.34) (5,3765.34) (7,376.34) (8,3765.34)};
            coordinates {(1,87.2) (2,86.0) (4,79.0) (5,73.6) (7,10.0) (8,18.6)};
        
        \addplot[fill=myGreen, fill opacity=0.75, text=ACMDarkBlue, pattern color=myGreen]
        	%coordinates {(1,7) (2,62.5) (4,47) (5,890) (7,3327) (8,16091.75)};
            coordinates {(1,1.6) (2,1.4) (4,9.9) (5,17.4) (7,88.8) (8,79.4)};

        % \addplot[fill=ACMDarkBlue, text=ACMDarkBlue, fill opacity=1]
        % 	coordinates {(1,4) (2,35) (3,269.6)};

            \node[] at (xticklabel cs:.2150,5pt) {1};
            \node[] at (xticklabel cs:.5000,5pt) {100};
            \node[] at (xticklabel cs:.7850,5pt) {10k};
    
            \node[] at (xticklabel cs:-0.05,5pt) {Batch Size:};
            \node[font=\small] at (xticklabel cs:-0.06,-5pt) {\% selectivity:};

        \legend{Normalization, Filtering, Computation}
        \end{axis}
        \end{tikzpicture}

    %\vspace*{-10pt}
    \caption{Vector dimensionality D=1024.}
    \label{fig:gemm_breakdown_1024}
    \vspace{-10pt}
    \end{subfigure}

    \caption{Tensor strategy time breakdown, varying the selectivity, dimensionality, and batch size.}
    \label{fig:gemm_breakdown_64_1024}
    %\vspace*{-10pt}
\end{figure}

\cref{fig:gemm_breakdown_64_1024} shows the breakdown of execution time across three dimensions: vector dimensionality (\cref{fig:gemm_breakdown_64}, \cref{fig:gemm_breakdown_1024}), batch size (1, 100, 10k), and selectivity ($10\%$,$90\%$). We normalize the times to $100\%$ to calculate the fraction of the particular execution strategy.
This validates the claim that where the computation dominates (large batch sizes), the Tensor-based approach is beneficial as it optimizes this phase. 
As a full filtering and data movement/copy step has to happen before (albeit linear in input size), it dominates the overall cost in both dimensionalities (but less so when $D=1024$). As relational filtering dominates, the more practical method is \texttt{Explicit SIMD} (\cref{fig:scan_comparison_dims}).
A mini-batch strategy similar to \textit{vectorized}\footnote{vectorization is becoming a severely overloaded term} data processing~\cite{DBLP:conf/cidr/BonczZN05} or sparse-vector computation could help the performance and overlap the filtering and computation steps, which remains a future research topic.

\paragraph{Takeaway}
Having two optimized scan-based strategies: individual hardware optimized for low batch sizes and a tensor-based strategy for large batches (or vector joins~\cite{DBLP:journals/corr/abs-2312-01476}) enables an optimizer (or a user) to select a more efficient workload-based method.

%%%%%%%%%%%%%%%%%%%%%%%%%%%%%%%%%%%%%%%%
\subsection{Selective queries: should I scan or probe?}

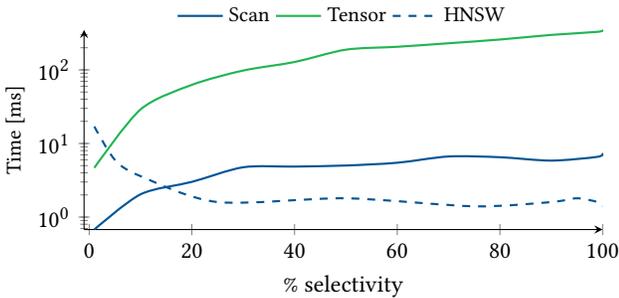
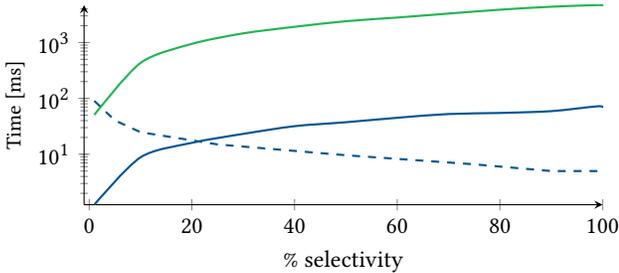
\begin{figure}
    %\color{ACMBlue} % for revision
    \centering
    
    \begin{subfigure}{\linewidth}
    \centering
    %%%%%%%%%%%%%%%%%%%%%%%%%%%%%%%%%%%%%%%%%%%%%%%%%%%%%%%%%%%%%
        \begin{tikzpicture}

        \begin{axis}[
            legend style={at={(0.5,1)},draw=none,
                anchor=south,
                legend columns=3,
                inner sep=0pt, outer sep=0pt, font=\small%,
            },
            axis lines = left,
            axis x line*=bottom,
            axis y line*=left,
            xlabel={\% selectivity},
            xlabel style={align=center},
            ylabel style={align=center},
            ylabel={Time [ms]},
            % ylabel near ticks,
            % legend style={font=\tiny},
            ylabel absolute, every axis y label/.append style={yshift=-1em, font=\small},
            height=0.5\columnwidth,
            width=\linewidth,
            xmin=-1,
            xmax=100,
            ymode=log,
            log basis y={10},
            % log ticks with fixed point,
            %line width=1.5pt
            %ymin=0,
            %xmin=1,
            %ymax=18000
        ]
            \addplot[smooth,color=ACMDarkBlue, thick]  coordinates {
                (1,0.6771235)
(10,2.037403)
(20,3.00941475)
(30,4.750953)
(40,4.853673)
(50,5.020636)
(60,5.47420225)
(70,6.658652)
(80,6.50607725)
(90,5.85161625)
(99,6.589551)
(100,7.40759775)
            };
            \addlegendentry{Scan}
    
        	\addplot[smooth,color=myGreen, thick]  coordinates {
        		(1,4.66666666666667)
(10,29)
(20,62.5833333333333)
(30,98)
(40,128)
(50,188.333333333333)
(60,206)
(70,230.083333333333)
(80,258.5)
(90,298.833333333333)
(99,328.916666666667)
(100,349.416666666667)
        	};
        	\addlegendentry{Tensor}

         \addplot[smooth,color=ACMDarkBlue, thick, dashed]  coordinates {
                (1,17)
                (5,6)
                (10,3.6)
                (25,1.6)
                (50,1.8)
                (75,1.4)
                (90,1.6)
                (95,1.8)
                (99,1.6)
                (100,1.4)
            };
            \addlegendentry{HNSW}

	\end{axis}
    \end{tikzpicture}

%\vspace*{-10pt}
\caption{Scan vs Probe D=64}
\label{fig:scan_vs_probe_vec_64_1}
%\vspace*{-10pt}
    \end{subfigure}
    
    \begin{subfigure}{\linewidth}
    \centering
    %%%%%%%%%%%%%%%%%%%%%%%%%%%%%%%%%%%%%%%%%%%%%%%%%%%%%%%%%%%%%
        \begin{tikzpicture}

        \begin{axis}[
            legend style={at={(0.5,1)},draw=none,
                anchor=south,
                legend columns=3,
                inner sep=0pt, outer sep=0pt, font=\small%,
            },
            axis lines = left,
            axis x line*=bottom,
            axis y line*=left,
            xlabel={\% selectivity},
            xlabel style={align=center},
            ylabel style={align=center},
            ylabel={Time [ms]},
            % ylabel near ticks,
            % legend style={font=\tiny},
            ylabel absolute, every axis y label/.append style={yshift=-1em, font=\small},
            height=0.5\columnwidth,
            width=\linewidth,
            xmin=-1,
            xmax=100,
            ymode=log,
            log basis y={10},
            % log ticks with fixed point,
            %line width=1.5pt
            %ymin=0,
            %xmin=1,
            %ymax=18000
        ]
           \addplot[smooth,color=ACMDarkBlue, thick]  coordinates {
                (1,1.24861325)
(10,8.8141985)
(20,16.01074175)
(30,23.15917075)
(40,31.8132565)
(50,37.44492)
(60,44.9744135)
(70,52.51678625)
(80,54.847327)
(90,59.08151375)
(99,72.188484)
(100,67.93502725)
            };
            %\addlegendentry{Scan}
    
        	\addplot[smooth,color=myGreen, thick]  coordinates {
        		(1,50.6666666666667)
(10,431.333333333333)
(20,945.5)
(30,1459.5)
(40,1912.58333333333)
(50,2408.33333333333)
(60,2802.91666666667)
(70,3302.5)
(80,3873.75)
(90,4377.83333333333)
(99,4683.33333333333)
(100,4598.25)
        	};
        	%\addlegendentry{Tensor}

         \addplot[color=ACMDarkBlue, thick, dashed]  coordinates {
                (1,90.2)
                (5,40.4)
                (10,24.8)
                (25,15)
                (50,9.6)
                (75,6.6)
                (90,5)
                (95,5)
                (99,5)
                (100,5.2)
            };
            %\addlegendentry{HNSW}

	\end{axis}
    \end{tikzpicture}

%\vspace*{-10pt}
\caption{Scan vs Probe D=1024}
\label{fig:scan_vs_probe_vec_1024_1}
\vspace*{-10pt}
    \end{subfigure}

    \caption{Different search strategies, query batch size is 1.}
    \label{fig:scan_vs_probe_vec_64_1024_1}
    %\vspace*{-10pt}
\end{figure}
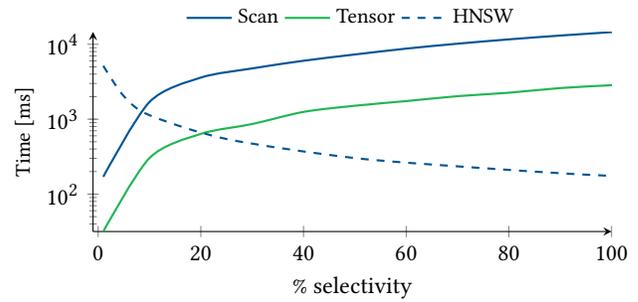
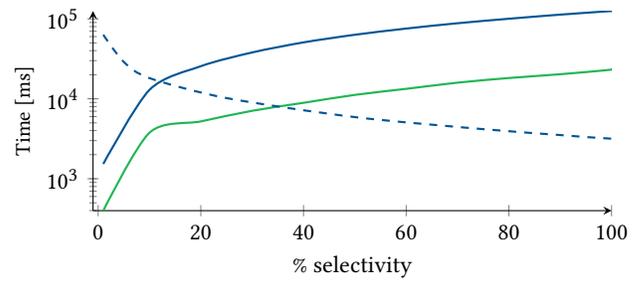
\begin{figure}
    %\color{ACMBlue} % for revision
    \centering
    
    \begin{subfigure}{\linewidth}
    \centering
    %%%%%%%%%%%%%%%%%%%%%%%%%%%%%%%%%%%%%%%%%%%%%%%%%%%%%%%%%%%%%
        \begin{tikzpicture}

        \begin{axis}[
            legend style={at={(0.5,1)},draw=none,
                anchor=south,
                legend columns=3,
                inner sep=0pt, outer sep=0pt, font=\small%,
            },
            axis lines = left,
            axis x line*=bottom,
            axis y line*=left,
            xlabel={\% selectivity},
            xlabel style={align=center},
            ylabel style={align=center},
            ylabel={Time [ms]},
            % ylabel near ticks,
            % legend style={font=\tiny},
            ylabel absolute, every axis y label/.append style={yshift=-1em, font=\small},
            height=0.5\columnwidth,
            width=\linewidth,
            xmin=-1,
            xmax=100,
            ymode=log,
            log basis y={10},
            % log ticks with fixed point,
            %line width=1.5pt
            %ymin=0,
            %xmin=1,
            %ymax=18000
        ]
            \addplot[smooth,color=ACMDarkBlue, thick]  coordinates {
(1,170.36730025)
(10,1699.50511625)
(20,3595.926672)
(30,4764.08666525)
(40,6032.0626555)
(50,7315.8424915)
(60,8760.86306875)
(70,10192.163048)
(80,11638.35950475)
(90,13089.33962775)
(99,14392.15184475)
(100,14650.86109575)
            };
            \addlegendentry{Scan}
    
        	\addplot[smooth,color=myGreen, thick]  coordinates {
(1,31.6666666666667)
(10,303.5)
(20,637.083333333333)
(30,862.5)
(40,1252.25)
(50,1513.33333333333)
(60,1745.25)
(70,2031.08333333333)
(80,2257.75)
(90,2606.08333333333)
(99,2820.91666666667)
(100,2855.66666666667)
        	};
        	\addlegendentry{Tensor}

         \addplot[smooth,color=ACMDarkBlue, thick, dashed]  coordinates {
(1,5168.2)
(5,2031)
(10,1131.6)
(25,544.4)
(50,302.8)
(75,222.2)
(90,190.2)
(95,181.8)
(99,176.4)
(100,171.8)
            };
            \addlegendentry{HNSW}

	\end{axis}
    \end{tikzpicture}

%\vspace*{-10pt}
\caption{Scan vs Probe D=64}
\label{fig:scan_vs_probe_vec_64_10k}
%\vspace*{-10pt}
    \end{subfigure}
    
    \begin{subfigure}{\linewidth}
    \centering
    %%%%%%%%%%%%%%%%%%%%%%%%%%%%%%%%%%%%%%%%%%%%%%%%%%%%%%%%%%%%%
        \begin{tikzpicture}

        \begin{axis}[
            legend style={at={(0.5,1)},draw=none,
                anchor=south,
                legend columns=3,
                inner sep=0pt, outer sep=0pt, font=\small%,
            },
            axis lines = left,
            axis x line*=bottom,
            axis y line*=left,
            xlabel={\% selectivity},
            xlabel style={align=center},
            ylabel style={align=center},
            ylabel={Time [ms]},
            % ylabel near ticks,
            % legend style={font=\tiny},
            ylabel absolute, every axis y label/.append style={yshift=-1em, font=\small},
            height=0.5\columnwidth,
            width=\linewidth,
            xmin=-1,
            xmax=100,
            ymode=log,
            log basis y={10},
            % log ticks with fixed point,
            %line width=1.5pt
            %ymin=0,
            %xmin=1,
            %ymax=18000
        ]
           \addplot[smooth,color=ACMDarkBlue, thick]  coordinates {
(1,1531.0772485)
(10,12943.79031975)
(20,25567.6368945)
(30,37999.59639475)
(40,50763.09983875)
(50,63232.40077675)
(60,75607.7368365)
(70,88125.162221)
(80,100303.86471775)
(90,112900.5498945)
(99,124062.42076225)
(100,125285.18099075)
            };
            %\addlegendentry{Scan}
    
        	\addplot[smooth,color=myGreen, thick]  coordinates {
(1,398.916666666667)
(10,3767.33333333333)
(20,5215.25)
(30,7081.5)
(40,8905.83333333333)
(50,11195.3333333333)
(60,13323.1666666667)
(70,15919.5)
(80,18187.5)
(90,20287.0833333333)
(99,22790.8333333333)
(100,23289)
        	};
        	%\addlegendentry{Tensor}

         \addplot[smooth,color=ACMDarkBlue, thick, dashed]  coordinates {
(1,63242.6)
(5,28891.4)
(10,18154.4)
(25,10212)
(50,5902)
(75,4171)
(90,3520.4)
(95,3325.2)
(99,3198.8)
(100,3171.8)
            };
            %\addlegendentry{HNSW}

	\end{axis}
    \end{tikzpicture}

%\vspace*{-10pt}
\caption{Scan vs Probe D=1024}
\label{fig:scan_vs_probe_vec_1024_10k}
\vspace*{-10pt}
    \end{subfigure}

    \caption{Different search strategies, query batch size is 10k.}
    \label{fig:scan_vs_probe_vec_64_1024_10k}
    %\vspace*{-10pt}
\end{figure}

Finally, we shed light on which access path to select: a scan-based one or an index-based one (probe). 
We extend the HNSW approach with filtering capabilities on relational columns by using a provided filtering function interface and probing the relational column using the unique vector ID. This imposes random access (per number of filters and qualifying elements) in addition to the vector index traversal. We use a filter on a single column in all of our experiments, controlling the relational selectivity.

At a lower selectivity, the index-based approach with pre-filtering will have to navigate more as the vectors might not satisfy the relational predicate, leading to more random accesses and vector comparisons (\cref{fig:scan_vs_probe_vec_64_1024_1}). 
As more values are selected, the approximate nearest neighbor algorithm can find sufficient tuples that satisfy the condition without excess comparisons.

While both scan-based strategies intersect with the index-based approach (\cref{fig:scan_vs_probe_vec_64_1}, \cref{fig:scan_vs_probe_vec_1024_1}), selecting an optimal strategy to improve the search performance is crucial. 
Results in \cref{fig:scan_vs_probe_vec_64_1024_10k} further reinforce the previous finding that the Tensor-based strategy is better for larger batch sizes, in both cases significantly improving the execution time until the intersection point with the parallel index-based approach, and resulting in an order of magnitude speedup if the most efficient of the methods is selected. 

\paragraph{Takeaway}
Selecting between scan-based and index-based access paths depends primarily on selectivity, dimensionality, and the batching strategy. It is worth noting that vector indexes can also be constructed with various parameters affecting the intersection point between the scan and the probe. Contrary to relational indexes, vector indexes perform better when more tuples satisfy the selection condition.

%%%%%%%%%%%%%%%%%%%%%%%%%%%%%%%%%%%%%%%%
\subsection{Looking Forward and Lessons Learned}
Selecting efficient access methods for mixed vector-relational search enriches the design space to improve end-to-end efficiency. It brings vector data management closer to traditional query optimization, especially if a vector index is not the main (vector) database structure~\cite{DBLP:conf/icde/SancaA23}. Use cases not purely a search, such as joins~\cite{DBLP:journals/corr/abs-2312-01476} that still rely on vector indexes can also provide workload-driven access path selection for more efficient execution, similar to prior art in purely relational operators.

{\small
\noindent
\begin{table}[h]
    \centering
    \begin{tabular}{c|c|c}
        & \textbf{Scan} & \textbf{Index}\\
        \hline
        \textbf{Accuracy} & Exact & Approximate \\
        %\hline
        %\textbf{Filtering} & Full Relational & Vector Similarity \& Pre-Filtering\\
        \hline
        \textbf{Cost} & Compute \& Scan & Build \& Compute \& Probe \\
        \hline
        \textbf{Flexibility} & Any Expression & Construction-Time Distance\\
        \hline
        \textbf{Benefit} & Selective Predicate & Full Search\\
        %\vspace{-5pt}
    \end{tabular}
    \caption{Index versus scan-based vector search.}
    %\vspace{-15pt}
    \label{tab:comparison}
\end{table}
}

We outline key differences between scan and index-based approaches in \cref{tab:comparison}.
Indexes are efficient but, at the same time, limited to the parameters given at construction time. If the query has different requirements, it is imperative to have an efficient fall-back option to an efficient exhaustive scan.
This might include queries over different vector distance functions than with which the index structure was constructed or range instead of top-k search for which an index might not be suitable (or the most efficient access path). 

From the modern hardware perspective, technologies such as High-Bandwidth Memory (HBM) or CPU-side accelerators for vector data processing such as the Intel Advanced-Matrix Extensions (AMX)~\cite{IntelAMX2023, DBLP:conf/vldb/SancaA23}, have the potential to change the computational and data access tradeoffs. Naturally, the GPU-side tensor processing units or using dedicated Tensor Processing Units (TPUs) further expands the domain of using accelerators or heterogeneous hardware, emphasizing the need for designing efficient and adaptive vector data management systems that efficiently use the underlying hardware       % 2 p       -- end of 21.03
\section{Conclusion}
\label{sec:conclusion}

Vector data management driven predominantly by index structures with support for efficient approximate nearest neighbor search faces a challenge once a single well-optimized workload assumption changes, especially when interacting with relational operators. For relational systems with support for vector indexes, selecting the most efficient data access path remains crucial for performance and efficiency. In this work, we presented and evaluated scan-based methods against an index-based search, and outlined the optimization space and parameters affecting the performance of data access paths for mixed vector-relational search, showing the trade-off between different methods and guidelines for selecting the most efficient and performant approach.

% Future work:
% \url{https://milvus.io/blog/deep-dive-1-milvus-architecture-overview.md#Data-Model}

% \url{https://milvus.io/blog/deep-dive-8-knowhere.md}

% data layouts for indexes (above reference shows that milvus is more of a row-store)
% test with annoy too: https://github.com/spotify/annoy
      % 0.25

%\begin{acks}
%\end{acks}

\bibliographystyle{ACM-Reference-Format}
\bibliography{sample-base}

\end{document}